\documentclass[pra,aps,superscriptaddress,balancelastpage,twocolumn]{revtex4}
\usepackage{amsmath}
\usepackage{graphicx}
\usepackage{amsfonts}
\usepackage{amssymb}

\providecommand{\U}[1]{\protect\rule{.1in}{.1in}}
\providecommand{\U}[1]{\protect\rule{.1in}{.1in}}

\begin{document}

\title{Light Drag in a Cavity Magnomechanics}
\author{Amjad Sohail}
\email{amjadsohail@gcuf.edu.pk}
\affiliation{Department of Physics, Government College University, Allama Iqbal Road,
Faisalabad 38000, Pakistan}
\affiliation{Instituto de Física Gleb Wataghin, Universidade Estadual de Campinas, Campinas, SP, Brazil}
\author{Hazrat Ali}
\email{yamanuom@gmail.com}
\affiliation{Department Physics, Abbottabad University of Science and Technology, Havellian, 22500, KPK, Pakistan}
\author{Khalid Naseer}
\affiliation{Department of Physics, University of Sargodha, Sargodha Pakistan}
\author{Rizwan Ahmed}
\affiliation{Physics Division, Pakistan Institute of Nuclear Science and Technology
(PINSTECH), P. O. Nilore, Islamabad 45650, Pakistan}

\begin{abstract}
The term "light dragging" describes how the trajectory of light changes as it travels through a moving medium. This phenomenon facilitates the precise detection of incredibly slow speeds of light, which is widely used in quantum gate operations, state transfer, and quantum memory implementations, etc. To the best of our knowledge, this is the first time we have proposed the existence of a light-dragging effect in a magnomechanical system (MMS). The origin of this crucial element stems from nonlinear dipole and magnetostrictive interactions in MMS. Magnomechanical characteristics such as magnon-photon and magnon-phonon couplings have a strong impact on both refractive and group index profile spectra. We also explore that lateral light drag shows a strong dependence on detuning by altering the amplitude and direction of the
translational velocity. This enabled us to alter the light's propagation within the magnomechanical system from superluminal to subluminal and vice versa by adjusting the probe's detuning. The ability to control and manipulate the light drag through the MMS could be helpful in designing novel devices with improved functionality at the microscopic scale.

\end{abstract}

\maketitle



\section{Introduction}
It has been well-known for a long time that light propagating through a moving medium shows a dragging effect along the direction of the medium. Historically, this effect was first theoretically proposed by Fresnel back in 1818 \cite{Fren}. Fresnel found that a light ray traveling at speed $v$ through a moving medium experiences lateral displacement $\Delta x =(n_{q}-n_{r}^{-1})(vL/c)$, where $v(c)$ is the speed of medium (light), $n_{g}(n_{r})$ is the group (phase) refractive index and $L$ in the length of the moving medium. Several years later, in 1851, Fizeau demonstrated this effect experimentally \cite{Fiz}. This dragging effect can be normal optical drag which is along the direction of the motion of the medium or anomalous optical drag which happens to be in the opposite direction of moving medium \cite{norm,opp}. In their studies, both Fresnel and Fizeau have ignored the dispersion effect of the refractive index. This discrepancy was later incorporated by Lorentz and Laub by considering the influence of dispersion on optical drag for a moving medium having fixed/moving boundaries in their independent studies \cite{lor,laub}. On the experimental side, many remarkable studies measured the dispersion effects on optical drag. These include the experiments by Zeeman and collaborators in various mediums like water, quartz, and flint glass \cite{zeem,zeem1,zeem2,zeem3,zeem4}. In later years, this research led to the advancement and understanding of Einstein's theory of special relativity. It is because of these observations, that Einstein assumed that the light-dragging effect does not occur and therefore, the speed of light is independent of its source's motion. It is also important to mention here that in a low dispersion regime, light dragging is negligibly small. However, to have observable results for a stronger drag effect, the moving medium must have either a long traveling distance or a large velocity. In the present era, there are many applications of optical drag effect. These include motion-sensor application \cite{norm}, light-drag velocimeter \cite{vel} and light speed reduction, etc \cite{reduc}.

To study and investigate the quantum effects at a macroscopic scale there are many platforms \cite{rev}. These include ultra-cold atoms/optical lattices \cite{cold,cold1,cold2}, quantum dots \cite{dot,dot1}, superconducting devices \cite{sup,sup1}, cavity optomechanical systems \cite{oms,oms1,oms2} and more recently the magnomechanical systems \cite{mag,mag1,mag2,mag3} etc. Out of these platforms, cavity magnomechanical systems provide a promising working platform in many aspects. These systems are mainly based on microwave (MW) field(s) of a cavity coupled to magnons associated with a single-crystal yttrium iron garnet ($Y_{3}Fe_{5}O_{12}$; YIG) sphere. In these systems, a magnetostrictive interaction acts like a radiation pressure analogous to the usual optomechanical system, and an applied magnetic field can drive the YIG sphere. In addition to magnetostrictive interaction, another interaction is known as magnetic dipole interaction \cite{mag,mag1}. It is because of the unique properties of the YIG sphere, that these systems offer a variety of remarkable features. Generally, YIG has a high Curie temperature, spin density, and small decay rates \cite{curie,curie1}. Countless interesting studies on the light-matter interaction cover a variety of quantum features on a macroscopic scale \cite{LM,LM0,LM1,LM2}. Nonetheless, light
drag has not yet been investigated in a magnomechanical system, although it might offer a vital role in studying the slow light phenomena.

In the present manuscript, we aim to study the optical drag effect in a magnomechanical system. To the best of our knowledge, this is one of the initial studies that incorporated the optical drag effect in a magnomechanical system. We consider a cavity magnomechanical system which is comprised of a single-mode microwave (MW) cavity having a YIG sphere driven by a uniform bias magnetic field (z-direction) which excites the magnon mode. The magnon mode and the cavity modes are coupled through magnetic-dipole interaction, whereas the phonon mode (due to the deformation of the YIG sphere under the action of incident radiation pressure) is coupled through magnetostrictive interaction. We consider a strong external MW field that drives the magnon mode of the YIG sphere. Our results show a strong dependence of optical drag upon the system parameters like normalized detuning, effective optomechanical coupling strength, and input power $P$.

The rest of the paper is organized as follows. In Section. 2, we present the model Hamiltonian of the magnomechanical system. We describe the dynamical equations of the magnomechanical system using the quantum Langevin approach. Furthermore, we derive the equation for the light drag by employing the refractive and group index. The analysis is reported in Section 3. Finally, we present the concluding remarks in Section 4.
\section{The Model}
We consider the standard magnomechanical system in which a YIG sphere is placed in a single-mode microwave (MW) cavity, as illustrated in Fig. 1.
The YIG sphere is subjected to a uniform bias magnetic field ($z$-direction),
which excites the magnon modes inside it. These modes are then coupled to
the cavity modes through magnetic-dipole interactions. Owing to fluctuating magnetization caused by the excitation of the magnon modes, the
lattice structure of the YIG spheres is deformed and, as a
result, the magnetostrictive interaction establishes the interactions
between the magnon and the phonon. The single-magnon magnomechanical coupling
strength depends on the diameter of the YIG sphere and the direction of the
external bias field and is very weak. However, we consider a strong
external microwave drive that drives the magnon mode of the YIG sphere. In
our model, this microwave drive acts as a control field and strengthens the magnon-phonon interaction inside the YIG sphere. Here, we consider a high-quality YIG sphere composed of ferric ions Fe$^{+3}$ of density $\rho
=4.22\times 10^{27}$m$^{-3}$ and diameter $D=250\mu $m. This results in a
total spin $S=\frac{5}{2}\rho V=7.07\times 10^{14}$m$^{-3}$, where $V$ is
the volume of the YIG sphere and $S$ denotes the collective spin operator.
The Hamiltonian of the system is given by

\begin{eqnarray}
\hat{H}/\hbar  &=&\omega _{c}c^{\dagger
}c+\omega _{m}m^{\dagger }m+\omega _{b}b^{\dagger }b  \notag \\
&&+\Gamma\left( cm^{\dagger }+c^{\dagger
}m\right) +g_{mb}m^{\dagger }m\left( b^{\dagger }+b\right)   \notag \\
&&+i\left( \varepsilon _{m}m^{\dagger }e^{-i\omega _{d}t}-\varepsilon
_{m}^{\ast }me^{i\omega _{d}t}\right)   \notag \\
&&+i\left( c^{\dagger }\varepsilon _{p}e^{-i\omega
_{p}t}+c\varepsilon _{p}^{\ast }e^{i\omega _{p}t}\right).
\end{eqnarray}%
The first three terms in Eq. (1) reflect the free Hamiltonian of the cavity
mode, magnon mode, and phonon mode. Here, $c_{k}^{\dagger }(c_{k})$, $%
m^{\dagger }(m)$, and $b^{\dagger }(b)$ are the creation (annihilation)
operators of the respective cavity mode, the magnon mode, and the phonon
mode, respectively. Furthermore, $\omega _{c}$, $\omega _{m}$, and $\omega
_{b}$, represent the respective resonance frequencies of the cavity, magnon,
and phonon modes. It is worth mentioning that the operators $m^{\dag }$ and
$m$ are the bosonic field operators for magnons and the frequency of magnon can be
determined by employing the gyromagnetic ratio $\gamma _{g}$ and the bias
magnetic field, $H$, via $\omega _{m}=\gamma _{g}H$. The fourth term
represents the interaction between the magnon modes and the cavity with
optomagnonical coupling strength $\Gamma$. The fifth term denotes the
interaction between the magnon and phonon modes with the magnomechanical coupling $%
g_{mb}$. The last three terms are input-driving field terms. The Rabi
frequency $\varepsilon _{m}=\frac{\sqrt{5N}}{4}\gamma _{g}H_{d}$ indicates
the strength of the coupling between the driving field of the microwave and
the magnon, where $N=\rho V$ stands for the YIG crystal's total spin number.
\begin{figure}[tbp]
\centering
\includegraphics[width=1\columnwidth,height=1.65in]{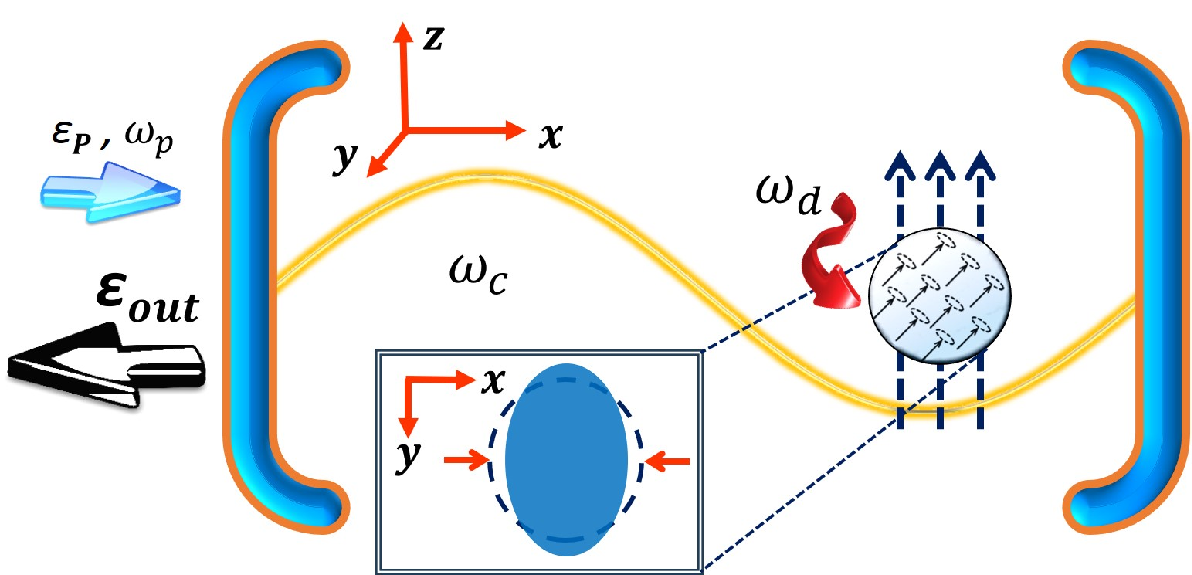} \centering
\caption{(a) Schematic diagram of a cross-line cavity magnoomechanical
system which coupled a magnon mode to both the cavity modes.}
\end{figure}
The total Hamiltonian for the current system about a frame rotating at the driving frequency $\omega _{d}$ is given by
\begin{eqnarray}
\hat{H}/\hbar  &=&\Delta _{c}c^{\dagger
}c+\Delta _{m}^{0}m^{\dagger }m+\omega _{b}b^{\dagger }b  \notag \\
&&+\Gamma\left( cm^{\dagger }+c^{\dagger
}m\right) +g_{mb}m^{\dagger }m\left( b^{\dagger }+b\right)   \notag \\
&&+i\varepsilon _{m}\left( m^{\dagger }-m\right) +i\left(c^{\dagger
}\varepsilon _{p}e^{-i\delta_{p} t}+c\varepsilon _{p}^{\ast }e^{i\delta_{p} t}\right).
\end{eqnarray}%
Here, $\Delta _{c}=\omega _{c}-\omega _{d}$ ($k=1,2$), $\Delta
_{m}^{0}=\omega _{m}-\omega _{d}$, and $\delta_{p}=\omega _{p}-\omega _{d}$ represent the frequency detunings of the cavity mode, the magnon mode, and the probe.

\subsection{Dynamics of the magnomechanical system}
To understand the dynamics of the system within the semiclassical
limit, we can write the Heisenberg-Langevin equations
\begin{eqnarray}
\dot{c} &=&-\left( i\Delta _{c}+\kappa _{c}\right) c-i\Gamma
m+\varepsilon _{p}e^{-i\delta_{p}t},  \notag \\
\dot{m} &=&-\left( i\Delta _{m}^{0}+\kappa _{m}\right)m -i\Gamma c-ig_{mb}m\left( b^{\dagger }+b\right)
+\varepsilon _{m},  \notag \\
\dot{b} &=&-(\gamma _{b}+i\omega _{b})b-ig_{mb}m^{\dagger }m.  \label{LAEQ}
\end{eqnarray}%
For the sake of clarity and simplicity and without the loss of generality, we have omitted the thermal and quantum input noise terms because we are
interested in investigating the mean response of the current system to the
applied probing field. Within the semiclassical perturbation framework, we
assume that the probe microwave field is substantially weaker than the
control microwave field. Consequently, we can expand each operator $z$ ($z=b,c,m$) as $%
z=z_{s}+\delta z$, where $z_{s}$ ($\delta z$) is the steady-state value
(small fluctuation) of the operator. First, we consider the steady-state
solutions, which are given by
\begin{eqnarray}
b_{s} &=&\frac{-ig_{mb}}{i\omega _{b}+\gamma _{b}}\left\vert
m_{s}\right\vert ^{2},  \notag \\
c_{s} &=&\frac{-i\Gamma m_{s}}{\kappa _{c}+i\Delta _{c}},  \notag
\\
m_{s} &=&\frac{\Omega \zeta _{c}}{\zeta _{c}\zeta
_{m}+\Gamma^{2}},  \label{MAV}
\end{eqnarray}%
where $\zeta _{s}=\kappa _{s}+i\Delta _{s}$ ($s=c,m$) and $\Delta
_{m}=\Delta _{m}^{0}+g_{mb}(b_{s}+b_{s}^{\ast })$. We assume that the current system is working in the resolved sideband regime, in which $\omega_{b}>>\kappa_m,\kappa_c$. In this regime, we can safely take $\Delta_{c}=\Delta_{m}=\omega_{b}$. Furthermore, Eq. (\ref{LAEQ}), can be easily solved by introducing slowly varying operators such as $\delta c=\delta ce^{-i\Delta _{c}t}$, $\delta m=\delta me^{-i\Delta _{m}}t$, and $\delta b=\delta be^{-i\omega _{b}t}$.
The amplitude of the probe field is assumed to be significantly weaker than the coupling of the external microwave drive on magnon mode. By taking
into account, the perturbation caused by the input probe field up to the
the first-order term, we obtain the set of linearized equations of motion
\begin{eqnarray}
\delta \dot{c} &=&-\kappa _{c}\delta c-\iota \Gamma \delta
m+\varepsilon _{p}e^{-i\delta_{p} }, \notag \\
\delta \dot{m} &=&-\kappa _{m}-\iota \Gamma \delta c-\iota G_{mb}\delta b, \notag \\
\delta \dot{b} &=&-\gamma _{b}\delta b-\iota G_{mb}^{\ast }\delta m,
\end{eqnarray}%
where $G_{mb}= g_{mb}m_s$ is the effective magnomechanical coupling coefficient. Note that for a fixed $g_{mb}$, the value of $G_{mb}$ can be modified/enhanced via $m_s$ by an external magnetic field (see Eq. (\ref{MAV})). In addition, we have assumed that $\sigma =\delta_{p}
-\omega _{b}$ is the effective detuning. To solve the above set of linearized
equations, we apply an ansatz $\delta z=z_{+}e^{-i\sigma}+z_{-}e^{i\sigma}$
where the coefficients $z_{+}$ and $z_{-}$ (with $z=c,m,b$), respectively, correspond to the components at the frequencies $\omega_p$ and $2\omega_d-\omega_p$.
Then it is straightforward to obtain the final solution at the probe frequency
\begin{eqnarray}
c_{+}&=&\frac{(\alpha _{m}\alpha _{b}+\left\vert G_{mb}\right\vert ^{2})\varepsilon _{p}}{\alpha _{c}(\alpha _{m}\alpha _{b}+\left\vert G_{mb}\right\vert ^{2}) +\Gamma ^{2}\alpha _{b}}, \label{OUT}
\end{eqnarray}%
where $\alpha _{z}=\kappa
_{z}-i\sigma $ ($z=c,m,b$). Based on the input-output theory $\varepsilon_{T}=\varepsilon_{in}-\kappa_{c}c$, we can write the equation for the amplitude of the output field at the probe frequency, given by
\begin{equation}
\varepsilon_{T}=\frac{2\kappa_c c_{+}}{\varepsilon _{p}}=\chi_{r}+i\chi_{i}.
\end{equation}
It is crucial to mention that $\varepsilon_{T}$ is a complex quantity. In addition, the real and imaginary parts of $\varepsilon_{T}$ exhibit the absorption (in-phase) and dispersion (out-of-phase) spectrum of the output field quadratures at the probe frequency.
\subsection{Light drag effect in magnomechanical system}
The novel idea is to discuss the light-dragging effects in MMS. The two main elements used to discuss the light-dragging effect are the refractive and group indices.
Since the output field is related to the optical susceptibility as $\chi=\varepsilon _{T}=\frac{2\kappa_c c_{+}}{\varepsilon _{p}}$, the refractive index of the output field at the probe frequency can be computed by $n_r=1+2\pi\chi$. Furthermore, the refractive index can be linked with the group index at the probe field in MMS as:
\begin{eqnarray}
n_{g}
&=&n_{r}+2\pi\omega \frac{\delta \chi}{\delta x},\nonumber \\
&=& 1+2\pi\chi+2\pi\omega \frac{\delta \chi}{\delta x}.
\end{eqnarray}
The output field comprises real and imaginary components; hence, the refractive and group indices of the MMS possess both real and imaginary parts, which are related to absorption and phase dispersion, respectively. Furthermore, group velocity, delay, advancement, and attenuation can be obtained from the group index of the system.  From another side, the lateral light drag in the optomechanical system can be written as
\begin{equation}
\Delta_{x}=(n_{g}-\frac{1}{n_{r}})\frac{v l}{c}. \label{LLD}
\end{equation}
The parameters $c$, $v$, and $l$ are the speed of light in vacuum, the translation velocity, and the length of the medium, respectively. Moreover, it can be seen from Eq. (\ref{LLD}), that the lateral light drag depends on both the group refractive index $n_g$ and the phase refractive index $n_r$.

\section{Discussion \label{secM}}
This section explicitly discusses the refractive index, group index, lateral light drag versus probe detuning and cavity translational velocity in the magnomechanical system
We vary the strength of magnon-photon coupling and input power of the magnon-phonon interaction and study its effect on the refractive index, group index, and lateral light drag.   We utilize the following parameters from a recent experiment on a hybrid magnomechanical system for numerical computation. $%
\omega_{c}=2\pi\times 10$ GHz, $\omega_{b}=2\pi \times 15$ MHz, $\kappa_{1}=2\pi \times 2.1$ MHz, $\kappa_{2}= 2\pi \times 0.15\kappa$, $\kappa_{m}=2\pi \times0.1$ MHz,
$\Gamma=2\pi \times 3.2$MHz, $\gamma_{b}=10^{-5}\omega_{b}$, $g_{mb}=2\pi \times 0.3$ Hz, $T=10$ mK, $H_d=1.3\times10^{-4}, \gamma_{G}/2\pi=28$GHz/T, $r=125\mu$m, and $\rho=4.22\times10^{27}m^{-3}$ \cite{phase}.

\begin{figure*}[tbp]
\centering
\includegraphics[width=1.98\columnwidth,height=5in]{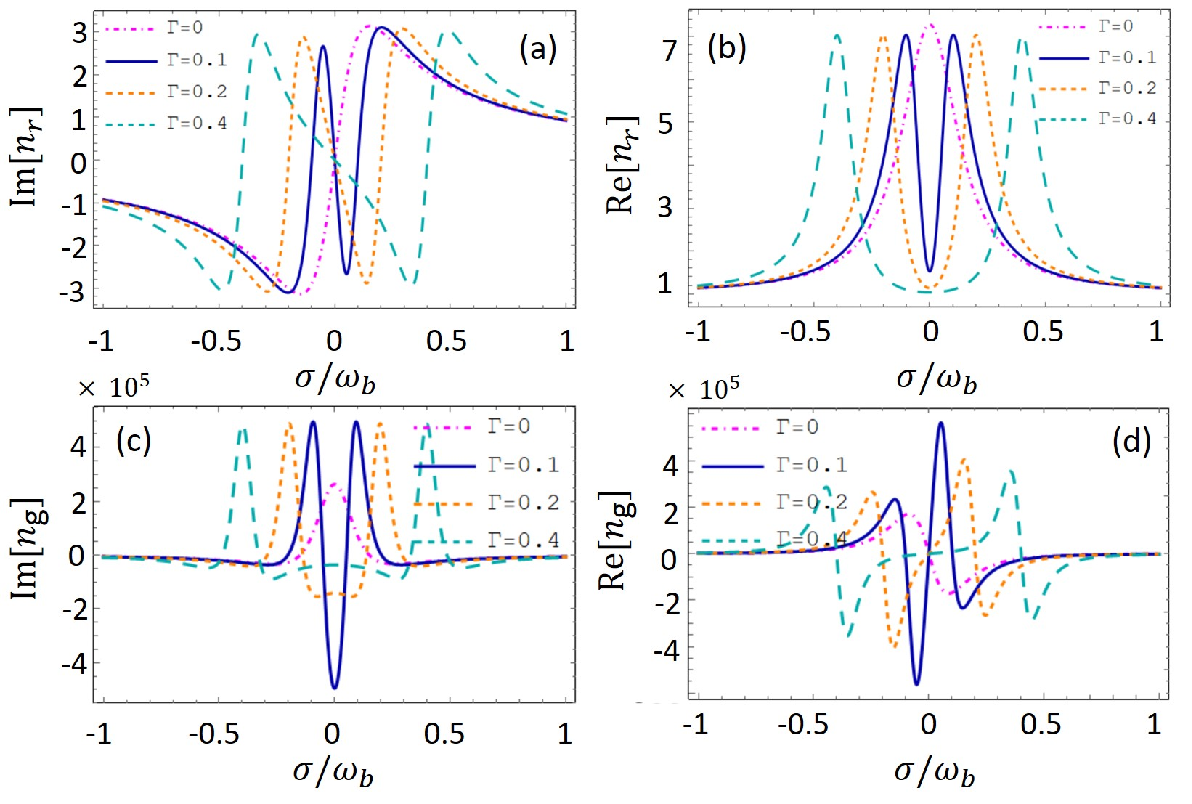} \centering
\caption{(a)(c) The imaginary and (b)(d) the real part of the (a)(b) refractive index and (c)(d) the group index as a function
of normalized detuning for different values of magnon-photon coupling.}
\end{figure*}
\begin{figure*}[tbp]
\centering
\includegraphics[width=1.98\columnwidth,height=2.7in]{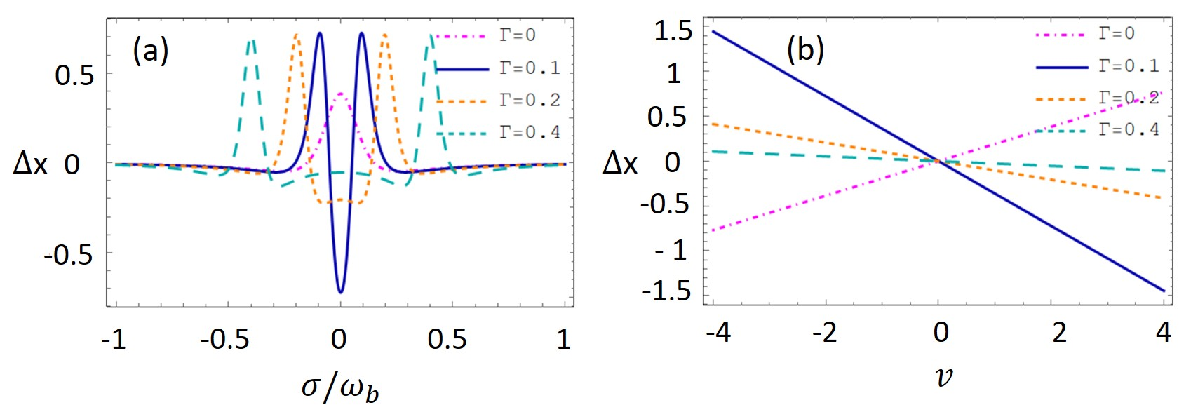} \centering
\caption{The light drag in an optomechanical system as a function of (a) x and (b) v for different values of magnon-photon coupling.}
\end{figure*}

Figure 2 shows the plot between the refractive index and group index versus normalized probe detuning by changing magnon-photon coupling strength while ignoring magnon-phonon interaction. Initially when the magnon-photon interaction is zero i.e., $\Gamma=0$, the slope of the Im $(n_{r}$) around the resonance is positive indicating the sub-luminal behavior of the light through the cavity see the pink curve of Fig 2 (a).  The slope of the dispersion spectrum around the resonance becomes negative (anomalous)  by considering the magnon-photon interaction in the cavity leads to super-luminal propagation of light around the resonance. Moreover, two normal dispersion slopes (at $\pm \sigma$) far from the resonance can be observed for each value of $\Gamma$ as depicted by the blue, red, and cyan curves of Fig 2 (a). The absorption spectrum ($Re (n_{r}$) is plotted against normalized probe detuning as shown in Fig 2 (b), The magnomechanical cavity completely absorbs the probe light when the magnon-photon interaction is not considered inside the cavity (see the pink curve of Fig 2 (b)). The cavity becomes completely transparent with two symmetrical absorption peaks for the prob light when the magnon-photon interaction is switched on. The width of the transparency window gets widened with increasing coupling strength of the magnon-photon interaction i.e., $\Gamma=0.1, 0.2, 0.4$, as elucidated by the blue, red, and cyan curves of Fig. 2 (b). Fig. 2 (c,d) is the group index of the opt-magnomechanical cavity versus probe detuning with changing the coupling strength of the magnon-photon interaction. The imaginary part of the group index of the cavity is positive around the resonance for zero coupling strength of magnon-photon interaction and we report $\pm 2.2 \times 10^{5}$, see pink curve of Fig 2 (c). The group index of the cavity is more sensitive for $\Gamma=0.1$ and we report both positive and negative of imaginary group index of about $\pm 4.5 \times 10^{5}$. Enhance normal and anomalous dispersion spectrum of the probe field through the cavity near the resonance ($\sigma=\pm 0.1$) is reported, see blue curve of Fig. 2 (c). The gain and absorption spectrum i.e., Re($n_{g})$ of the magnomechanical cavity is plotted against normalized probe detuning in Fig 2 (d). We observe absorption for negative detuning and gain for positive detuning through the cavity in the absence of magnon-photon interaction (pink curve). The absorption changes to gain spectrum and gain to absorption of the probe light through the cavity in the presence of magnon-photon interaction as elucidated by blue, red, and cyan curves of Fig 2 (d). The cavity response to the probe is highly sensitive for $\Gamma=0.1$, where an enhanced gain and absorption spectrum is achieved. Thus this particular value of the coupling strength of magnon-photon interaction can be used to achieve an intense laser beam through the cavity. Moreover, we observe PT-symmetric like behavior as the gain and absorption balance each other. Additionally, the quasi PT-symmetric behavior can be observed for both positive and negative detuning in the presence of strong magnon-photon interaction, which has technological application in optics and sensors.

The lateral light darg through the magnomechanical cavity versus probe detuning and translational velocity of the cavity in the presence of magnon-photon interaction is studied and the results are shown in Fig. 3 while keeping no input power of the magnon-phonon interaction is considered in the cavity. We observe a positive shift of light drag through the cavity when no magnon-photon interaction is considered see pink curve of figure 3 (a). We observe positive and negative shifts of light of $\pm 0.7 cm$ through the cavity for $\Gamma=0.1$. Increasing the coupling strength of magnon-photon interaction led us to enhance almost positive light drag see red and cyan curves of Fig 3 (a). To study the effect of translational motion on the light drag through the cavity, we plot $\Delta x$ versus velocity $v$ as shown in Fig 3 (b). The shift is along the direction of velocity, that is, the positive shift is for positive velocity, and the negative shift of light appears for negative velocity in the absence of magnon-photon interaction (as shown by the pink curve). The slope is positive, leading to subluminal propagation of light through the cavity. When the magnon-photon interaction is considered in the system, the behavior of the light drag becomes opposite i.e., the positive shift of light appears in the opposite direction to the translational velocity of the cavity leading to super-luminal probe propagation through the cavity. Moreover, the light drag is more sensitive for $\Gamma=0.1$ and we observe enhanced results up to $\pm 1.5 cm$ of the light drag.

\begin{figure*}[tbp]
\centering
\includegraphics[width=1.98\columnwidth,height=5in]{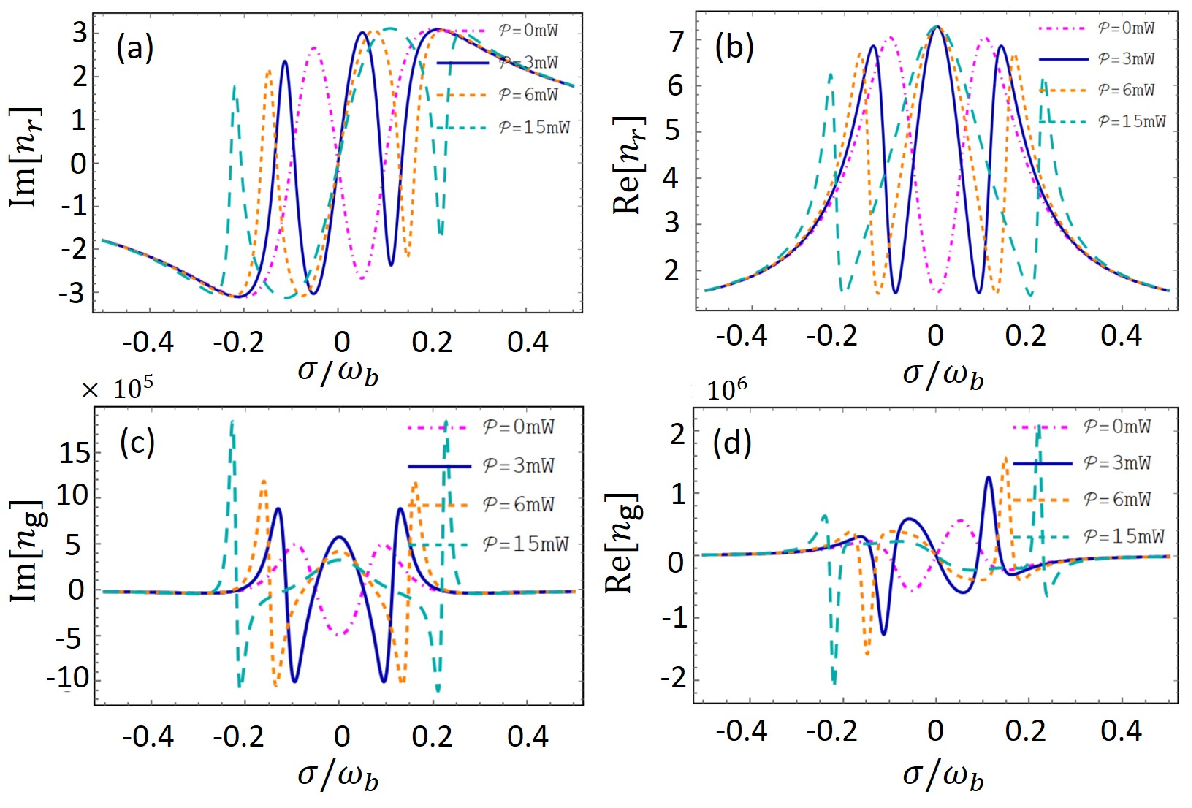} \centering
\caption{(a)(c) The imaginary and (b)(d) the real part of the (a)(b) refractive index and (c)(d) the group index as a function
of normalized detuning for different value of input power.}
\end{figure*}

\begin{figure*}[tbp]
\centering
\includegraphics[width=1.98\columnwidth,height=2.7in]{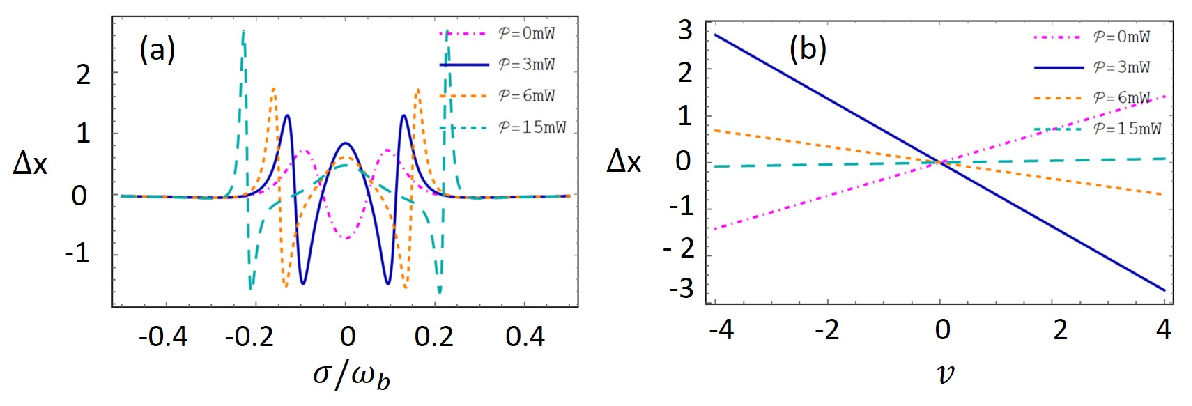} \centering
\caption{The light drag in an optomechanical system as a function of (a) $\sigma/\omega_{b}$ and (b) $v$ for different value of input power.}
\end{figure*}
\begin{figure*}[tbp]
\centering
\includegraphics[width=1.98\columnwidth,height=5in]{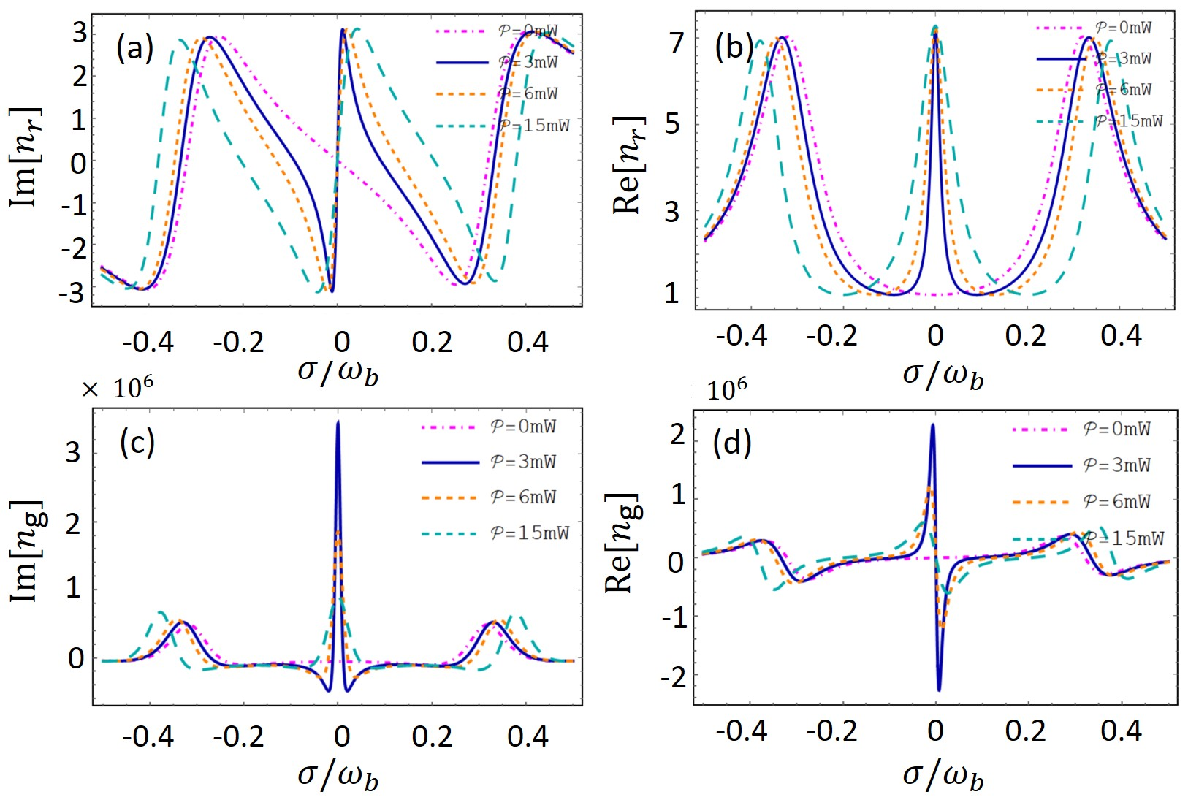} \centering
\caption{(a)(c) The imaginary and (b)(d) the real part of the (a)(b) refractive index and (c)(d) the group index as a function
of normalized detuning for different values of input power.}
\end{figure*}
To present more fascinating results of refractive index and group index of the magnomechanical cavity by keeping low magnon-photon interaction ($\Gamma=0.1$) constant and varying input power of magnon-phonon interaction, we plot $n_{r}$ and  $n_{g}$ versus normalized probe detuning as shown in Fig. 4. Initially when the input power of magnon-phonon interaction is zero, we observe anomalous dispersion curve around the resonance and two normal dispersion curves far from resonance as elucidated by the pink curve of Fig. 4(a). The anomalous dispersion curve of the probe light through the cavity changes to a normal dispersion curve around the resonance, when the input power of the magnon-phonon interaction is considered in the system, see blue, red, and pink curves of Fig 4 (a). Additionally, we observe two anomalous dispersion curves at ($\sigma = \pm 0.1$) and two normal dispersion curves at ($\sigma = \pm 0.15$) when the input power is 3 mw, as shown by the blue curve of Fig 4 (a). Thus by increasing the input power of the magnon-phonon interaction, we observe a similar pattern at larger detunings. The sub and super-luminal propagation of light through the cavity can be controlled at the desired detuning by varying the input power of magnon-phonon interaction.  Fig 4 (b) presents the absorption spectrum of the probe field through the magnomechanical cavity against normalized probe detuning. We observe OMIT around the probe resonance through the cavity when no input power of magnon-phonon interaction is applied, as depicted by the pink curve. We notice three absorption peaks and two OMIT windows in the presence of the input power of magnon-photon interaction, as shown by the blue, red, and cyan curves. The two transparency widows widen with the magnon-phonon interaction's increasing power; see the red and cyan curves. Moreover, the probe losses reduce at far resonance when the input power of the magnon-phonon increases, see the cyan-colored absorption peaks at $\sigma = \pm 0.25$. Figures 4 (c,d) demonstrate the imaginary and real of group index against normalized probe detuning in the presence of input power of magnon-phonon interaction. We observe anomalous and normal dispersion curves near the resonance when $\wp = 0$ as depicted by the pink curve of Fig 4 (c). The imaginary of group index of the cavity experiences opposite behavior near the resonance with the increasing input power. We also observe two additional dispersion (anomalous and normal) curves at $\sigma= \pm 0.1, \pm 0.15, \pm 0.2$ for varying power of magnon-phonon interaction see blue, red, and cyan of Figure 4 (c). Moreover, the dispersion curve gets steeper with increasing input power see the cyan curve for $\wp=15 mw$. 
We observed a three (3) fold enhancement if we only consider magnon-phonon interaction. The gain and attenuation profile of the cavity in the presence of both magnon-photon and magnon-phonon interactions are presented in Fig 4 (d). We observe a very low gain and absorption of the cavity by considering $\wp = 0$. We notice the gain of the probe light at negative detunings and absorption at positive detuning regions with the increasing value of p. A gradual increase in the gain spectrum is observed with a gradual increase of the input power of the magnon-phonon interaction. Thus a more intense laser beam can be obtained by considering strong magnon-phonon interaction. The PT-symmetric like behavior is obvious as the loss and gain balance each other.
\begin{figure*}[tbp]
\centering
\includegraphics[width=1.98\columnwidth,height=2.6in]{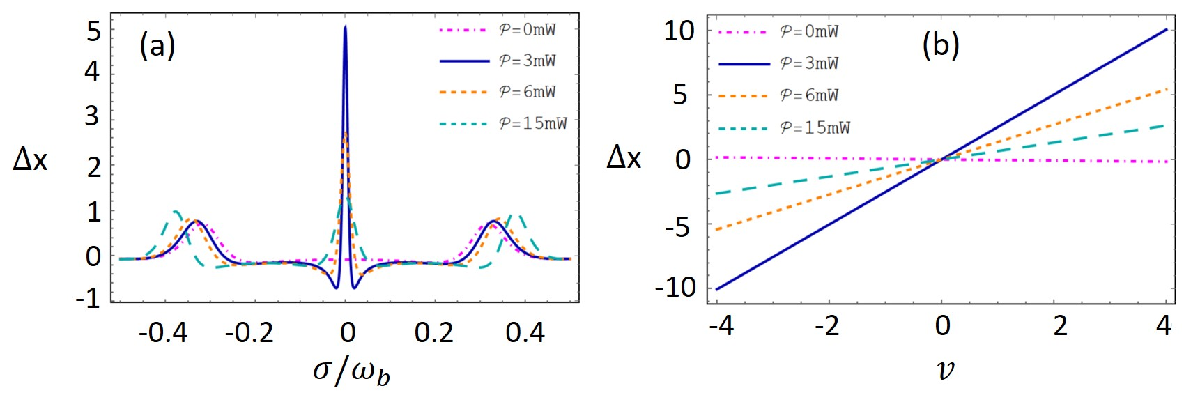} \centering
\caption{The light drag in an optomechanical system as a function of (a) $\sigma/\omega_{b}$ and (b) $v$ for different values of input power.}
\end{figure*}

To study the effects of light drag through the cavity against normalized probe detuning and cavity translational velocity for keeping constant low magnon-photon interaction ($\Gamma = 0.1$) and varying the input power of magnon-phonon interaction ($\wp$), we plot $\Delta x$ versus $\sigma$ and $v$, as shown in Fig. 5. We observe a negative light drag of $0.8 cm$ through the cavity around the resonance and a positive light drag of $0.6 cm$ at $\sigma = \pm 0.1$ when $\wp = 0$, as shown by the pink of Fig. 5 (a). Two positive peaks of light darg of $1.2 cm$ at $\sigma = \pm 0.1$ and two negative peaks of light drag of $1.4 cm$ at $\sigma = \pm 0.15$ are observed as the input power of $\wp= 3$mw is applied in the cavity, see the blue curve of figure 5 (a). As the power of the magnon-phonon is further increased to $6$mw, we observe enhanced positive and negative peaks of light drag of +1.7 cm and -1.5 cm, respectively, as depicted by the red curve. Further enhancement of of light darg of +2.7 and -1.6 through the cavity is observed, as the input power is increased to $15$mw, as shown by the cyan curve. Thus 4.5 times enhanced positive light drag and 2 times negative light darg is achieved for strong magnon-phonon interaction. Figure 5 (b) shows the light drag through the cavity versus its moving velocity for constant weak magnon-photon interaction and varying power of the magnon-phonon interaction at probe resonance. Initially, when $p=0$, we observe a positive slope indicating sub-luminal probe propagation through the cavity see the pink curve. The slope changes to negative as the power of the magnon-phonon interaction increases showing the super-luminal propagation through the cavity. The light drag $\Delta x$ of $\pm 3$ cm is observed for $\wp=3$mw, which is almost twice that of Figure 3 (b).

We consider the strong interaction of magnon-photon interaction ($\Gamma=0.4$) and varying the power of magnon-phonon interaction and investigate the refractive index and group index of the cavity in Figure 6. Initially, when the magnon-phonon interaction is zero ($\wp=0$), we observe an anomalous dispersion curve (super-luminal propagation) around the probe resonance, see the pink curve of Figure 6 (a). The anomalous dispersion curve changes to a very sharp normal dispersion curve around the resonance leading to slow light traveling in the cavity as depicted by the blue, red, and cyan curves of Figure 6 (a). Figure 6(b) shows Re ($n_{r}$), the absorption profile of the light through the cavity. The cavity is completely transparent with only one transparency window in the absence of magnon-phonon interaction ($\wp=0$). We notice two transparency windows of the probe through the cavity as magnon-phonon interaction is considered [$\wp=(3, 6, 15)$mw]. The two transparent windows of the cavity further widen for the higher power of magnon-phonon interaction see cyan curve of Figure 6 (b). The imaginary of $n_{g}$ is plotted versus normalized detuning in Figure 6(c). We notice anomalous and normal sharp dispersion slopes for a minimal range of probe detuning around the resonance. Furthermore, we obtain anomalous dispersions at far resonance for negative detuning and normal dispersion slopes at far resonance on positive detuning. Figure 6 (d) is the Re $n_{g}$ (gain and absorption profile) for the strong interaction of magnon and photon while changing input power magnon-phonon interaction. We notice a nearly flat line of the real of group index around the resonance when $\wp=0$. A sharp absorption and gain peaks are achieved by considering the magnon-phonon interaction. The gain and absorption through the cavity are very sensitive for $\wp = 3$mw. A more intense laser beam through the cavity is obtained on the probe resonance.

Finally, we investigate the light drag in the presence of strong magnon-photon interaction and the changing input power of magnon-phonon interaction. We notice a 0.5cm shift of the light through the cavity around the resonance and 1cm at both $\pm \sigma$ when $\wp=0$, see pink curve of Fig. 7 (a). The light drag of $-0.8 cm$ and $5 cm$ through the cavity is observed near the resonance when the input power of $3$mw is applied, as depicted by the blue curve. Thus 8.25 enhanced positive light darg is observed when considering both the strong magnon-photon interaction and magnon-phonon interaction. We notice the slope of the light drag from positive to negative through the cavity indicating super-luminal propagation when $p=0$, see the pink curve of Fig. 7(b). The slope of the light drag changes from negative to positive through the cavity showing sub-luminal behavior of light when input power is applied in the system. We notice $pm 10 cm $ of light drag through the cavity when $p=3$, which is 3.3 times an enhanced result of light drag, compared to Fig. 5(b).
\section{Conclusions}
The present study demonstrated that magnetostrictive and magnetic dipole interactions can undergo light drag. Consequently, we examined the refractive and group indices, both associated with light propagation. The effect of magnon-photon interaction and magnon-phonon power on the refractive index, group index, and light drag versus probe detuning and translational velocity of the cavity through the magnomechanical system is investigated.  We observed substantial changes in refractive index and group index when only the magnon-photon interaction is considered. Moreover, we noticed the negative group index in the negative probe detuning region, while the positive group index for the positive detuning region showed gain and absorption of the light through the cavity, simultaneously. The slope of the light drag is changed from positive to negative when the magnon-photon interaction is considered in the medium, leading to super-luminal propagation of light through the cavity.  An enhanced gain for negative detuning and absorption for positive detuning are observed by considering the input power of the magnon-phonon, while the magnon-photon interaction is kept low in the cavity. A maximum of $\pm 3$ cm of light drag is observed for both the low-magnon-photon interaction and the low power of the magnon-phonon interaction. However, we obtained a maximum of $\pm 10$ cm of light drag for strong magnon-photon and magnon-phonon interaction.
The light drag in a magnomechanical system will not only facilitate the advancement of next-generation photonic devices, but also indicate potential applications in the sensitivity of magnomechanical systems. Consequently, we assert that our technique possesses the capability to be employed with the present-day technology in quantum information processing.


\begin{thebibliography}{99}

\bibitem{Fren} A. J. Fresnel, Letter from Mr. Fresnel to Mr. Arago, on the
influence of earthly movement in some optical phenomena,
Ann. Chim. Phys. 9, 57 (1818).

\bibitem{Fiz} M. H. Fizeau, Hypotheses relating to the luminous ether and on an experiment which appears to demonstrate that the motion of bodies changes the speed with which light propagates in their interior, C. R. Acad. Sci. 33, 349 (1851).

\bibitem{norm} Qin T, Yang J, Zhang F, Chen Y, Shen D, Liu W, Chen L, Jiang X, Chen X and Wan W, Fast-and slow-light-enhanced light drag in a moving microcavity, Commun. Phys. 3 118 (2020).

\bibitem{opp} Banerjee C, Solomons Y, Black A N, Marcucci G, Eger D, Davidson N, Firstenberg O and Boyd R W, Anomalous optical drag, Phys. Rev. research 4 033124 (2022).

\bibitem{lor} H. A. Lorentz, Electromagnetic phenomena in a system moving with any velocity smaller than that of light, Proc. KNAW 6, 809 (1904).

\bibitem{laub}  J. Laub, To the optics of moving bodies, Ann. Phys. (NY) 330, 175 (1908).

\bibitem{zeem} P. Zeeman, Fresnel’s coefficient for light of different colors (first
part), Proc. KNAW 17, 445 (1914).

\bibitem{zeem1} P. Zeeman, Fresnel’s coefficient for light of different colors (second part), Proc. KNAW 18, 398 (1915).

\bibitem{zeem2} P. Zeeman, The propagation of light in moving transparent solid substances. I. Apparatus for the observation of the Fizeau-effect in solid substances, Proc. KNAW 22, 462 (1919).

\bibitem{zeem3} A. Snethlage and P. Zeeman, The propagation of light in moving, transparent, solid substances. II. Measurement of the Fizeau effect in quartz, Proc. KNAW 22, 512 (1920).

\bibitem{zeem4} P. Zeeman, W. D. Groot, A. Snethlage, and G. C. Dibbetz, The propagation of light in moving, transparent, solid substances. III. Measurements on the Fizeau effect in flint glass, Proc.
KNAW 23, 1402 (1922).

\bibitem{vel} Z. Chen, H. M. Lim, C. Huang, R. Dumke, and S.-Y. Lan, Quantum-Enhanced Velocimetry with Doppler-Broadened Atomic Vapor, Phys. Rev. Lett. 124, 093202 (2020).

\bibitem{reduc} Hau, L. V., Harris, S. E., Dutton, Z. and Behroozi, C. H. Light speed reduction to
17 metres per second in an ultracold atomic gas. Nature 397, 594–598 (1999).

\bibitem{rev} F. Fr\"{o}wis, P. Sekatski, W. D\"{u}r, N. Gisin, and N. Sangouard,
Macroscopic quantum states: Measures, fragility, and implementations, Rev. Mod. Phys. 90, 025004 (2018).

\bibitem{cold} Helmut Ritsch, Peter Domokos, Ferdinand Brennecke and Tilman Esslinger, "Cold atoms in cavity-generated dynamical optical potentials", Rev. Mod. Phys. 85, 553 (2013).

\bibitem{cold1} Greiner, M., Fölling, S. Optical lattices. Nature 453, 736–738 (2008).

\bibitem{cold2} Schäfer, F., Fukuhara, T., Sugawa, S. et al. Tools for quantum simulation with ultracold atoms in optical lattices. Nat Rev Phys 2, 411–425 (2020).

\bibitem{dot} Zhan, Y., Sun, S. A single quantum dot passively mediates entanglement. Nat. Phys. 20, 1363–1364 (2024).

\bibitem{dot1} D. Vashaee and J. Abouie, "Microwave-induced cooling in double quantum dots: Achieving millikelvin temperatures to reduce thermal noise around spin qubits
", Phys. Rev. B 111, 014305 (2025).

\bibitem{sup} Muhammad Waseem, Rizwan Ahmed, Muhammad Irfan and Shahid Qamar, “Three qubit Grover’s algorithm using superconducting quantum interference devices in cavityQED”, Quant. Inf. Proc. 12, 3649 (2013).

\bibitem{sup1} Cleidson Castro, Matheus R Araújo and Clebson Cruz, "Entanglement dynamics of a dc SQUID interacting with a single-mode radiation field", Phys. Scr. 96 105101 (2021).

\bibitem{oms} Markus Aspelmeyer, Tobias J. Kippenberg and Florian Marquardt, "Cavity optomechanics", Rev. Mod. Phys. 86, 1391 (2014).

\bibitem{oms1} Amjad Sohail, Rizwan Ahmed and Chang shui Yu, “Switchable and Enhanced Absorption
via Qubit-Mechanical Nonlinear Interaction in a Hybrid Optomechanical System”, Int. J.
Theor. Phys. 60, 739 (2021).

\bibitem{oms2} Rizwan Ahmed and Shahid Qamar, “Optomechanical entanglement via nondegenerate
parametric interactions”, Phys. Scr. 92, 105101 (2017).

\bibitem{mag} Jie Li, Shi-Yao Zhu and G.S. Agarwal, Magnon-Photon-Phonon Entanglement in Cavity Magnomechanics, Phys. Rev. Lett. 121, 203601 (2018).

\bibitem{mag1} Xuan Zuo, Zhi-Yuan Fan, Hang Qian, Ming-Song Ding, Huatang Tan, Hao Xiong, Jie Li, "Cavity magnomechanics: from classical to quantum", New J.Phys. 26 031201 (2024).

\bibitem{mag2} Amjad Sohail, Rizwan Ahmed, Rida Zainab and Chang- shui Yu, Enhanced entanglement and quantum steering of directly and indirectly coupled modes in a magnomechanical system,  Phys. Scr. 97 075102 (2022).

\bibitem{mag3} Amjad Sohail, Rizwan Ahmed, Jia-Xin Peng, Aamir Shahzad, and S. K. Singh, Enhanced entanglement via magnon squeezing in a two-cavity magnomechanical system, J. Opt. Soc. Am. B 40(5) 1359-1366 (2023).

\bibitem{curie} D. Zhang, X. M. Wang, T. F. Li et al., “Cavity quantum electrodynamics with
ferromagnetic magnons in a small yttrium-iron-garnet sphere,” npj Quantum
Inf. 1, 15014 (2015).

\bibitem{curie1} Y. Liu, L. Ling, T. Shui, N. Ji, S. Liu, and W.-X. Yang, “Two-color second-order
sideband generation via magnon Kerr nonlinearity in a cavity magnonical
system,” J. Opt. Soc. Am. B 39, 1042–1049 (2022).


\bibitem{LM} Ming-Song Ding, Li Zheng and Chong Li, Phonon laser in a cavity magnomechanical system
, Scientific Reports 9 15723 (2019).


\bibitem{LM0} A. Sohail, M Qasymeh, and H. Eleuch, Entanglement and quantum steering in a hybrid quadpartite system. Phy. Rev. APPLIED 20, 054062 (2023).

\bibitem{LM1} Amjad Sohail, Rizwan Ahmed, Aamir Shahzad and Muhammad Aslam Khan, “Magnon-Phonon-Photon Entanglement via the Magnetoelastic Coupling in a Magnomechanical System”, Int. J. Theor. Phys. 61, 174 (2022).

\bibitem{LM2} Xufeng Zhang, Chang-Ling Zou, Liang Jiang and Hong X. Tang, Cavity magnomechanics

\bibitem{phase} Xiyun Li, Wen-Xing Yang, Tao Shui, Ling Li, Xin Wang and Zhen Wu, Phase control of the transmission in cavity magnomechanical system with magnon driving,  J. Appl. Phys. 128, 233101 (2020).
\end{thebibliography}
\end{document}